\documentstyle [pre,aps,epsfig,multicol]{revtex}
\newenvironment{mytitle}{\begin{center} \large \bf }{\\ [.1in]\end{center}}
\newenvironment{myauthor}{\begin{center} \large }{\\ [.1in]\end{center}} 
\newenvironment{myinstit}{\begin{center} \large \it}{\end{center}}
\begin{document}

\thispagestyle{empty}

\begin{mytitle}
{\LARGE \bf Directional Emission from Microlasers with Open Chaotic Resonators}
\end{mytitle}

\begin{myauthor}
J. A. M\'endez-Berm\'udez$^1$, G. A. Luna-Acosta$^1$, and P. \v{S}eba$^{2,3}$
\end{myauthor}

\begin{myinstit}
$^1$Instituto de F\'{\i}sica, Universidad Aut\'onoma de Puebla, Apartado Postal J-48, Puebla 72570, M\'exico\\
$^2$Department of Physics, University Hradec Kralove, Hradec Kralove, Czech Republic\\
$^3$Institute of Physics, Czech Academy of Sciences, Cukrovarnicka 10, 
Prague, Czech Republic\\
\end{myinstit}

\begin{center}
{\bf Abstract} 
\end{center}
{\small
Using topological classical considerations (Poincar\`{e} Maps) the appearance of ``bow-tie''-shaped resonances is predicted for certain types of waveguides. The geometry of such waveguides is chosen to yield mixed chaotic dymnamics. Then, based on resonance effects, we propose the construction of directional emission microlasers using {\it open} chaotic semiconductor resonators of high refractive index.\\

}

\section{Introduction}

Due to the large variety of technological applications, semiconductor lasers have become the most important class of lasers \cite{appl}. In particular, the characteristics of the resonator, which is an essential component of lasers, play a fundamental role in the performance and size of these devices. This justifies the effort paid in the improvement of resonators. One example is the recent proposal and development of the {\it asymetric resonant cavity} (ARC) as resonator for semiconductor lasers \cite{arc1,arc2}. The ARC's are 2D resonators deformed from circular symmetry. If the deformation is weak {\it whispering-gallery}-type modes are present and produce highly directional emmission \cite{arc1}. While for large enough deformations, {\it bow-tie}-shaped resonances are responsible for the improvement in the power and directionality of the laser emmission \cite{arc2}.\

Here we propose the use of a wave-guide system as an {\it open} resonator, which for certain geometries produce {bow-tie}-shaped resonances.

\section{The System}

As a model of open resonator we use a 2D wave-guide which consists of two semi-infinite leads of width $d$, both along the $x$-axis, conected to a cavity with two walls: a wall described by the function $y(x)=d+a \xi(x)$, and a flat wall by $y=0$; $a$ is the amplitude of the deformation and we choose $\xi(x) = 1-\cos(2\pi x/L)$. Note that $\xi(x)$ is a periodic function of $x$, but in this work we are using only one period of it to construct the cavity, {\it i.e.} the length of the cavity is equal to $L$.\

We consider to be in the limit of ray optics and assume that our wave-guide system is made of a semiconductor with refractive index $n$. Also, we assume to have a metalic layer at $y=0$ and $y=d$ on the leads and at the $y=0$ boundary of the cavity, so that, light inside the wave-guide can only scape from it trough the deformed boundary of the cavity.\

We start the analysis of the wave-guide system by reviewing the classical motion in the infinitely long periodic cavity \cite{infinite}. To get the dynamical panorama, we look at a Poincar\`{e} Map (PM) of the system. The dynamics can be regular, mixed, or fully chaotic depending on the geometrical parameters. In particular, for $(d,a,L) = (1,0.305,5.55)$ the dynamics is mixed \cite{finite1}. These are the parameters that specify the shape of the cavity in this paper.\

Since we are interested in the dynamics of rays impinging on the boundary at $y(x)=d+a \xi(x)$ (because for certain incidence angles $\chi$ these rays can scape from the wave-guide) we locate the Poincar\`{e} Surface of Section (PSS) on this wall to construct the PM. Moreover, we plot the sinus of the angle of incidence of a ray to the boundary $\sin(\chi)$ and the coordinate $x$, instead of the traditional Birkhoff variables \cite{infinite,finite1}. In Fig. 1 we present the PM which shows three islands; one located at $\sin(\chi)=0$, and the other two centered at $\sin(\chi)=0.3$.

\begin{figure}[htb]
\begin{center}
\epsfig{file=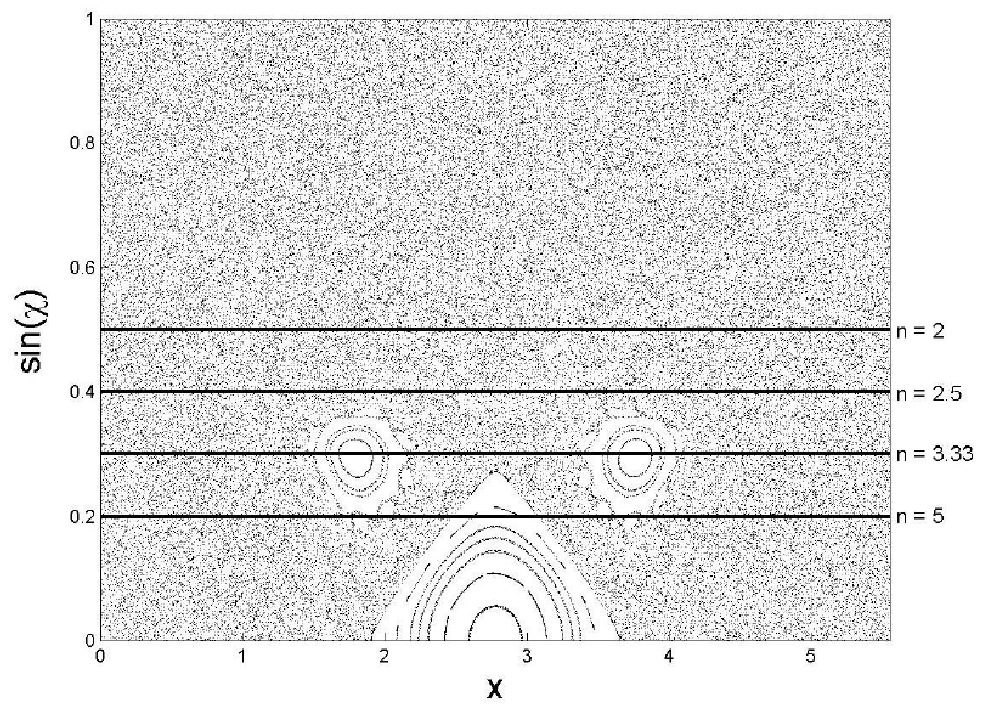,width=5in,height=3.5in}
\label{weaal0}
\end{center}
\end{figure}
\vspace{-0.28in}
{\small FIG. 1 Poincar\`{e} Map for the periodic cavity. The Poincar\`{e} Surface of Section is located at $y(x)=d+a [1-\cos(2\pi x/L)]$ with $(d,a,L) = (1,0.305,5.55)$. Critical lines for refractive scape are shown for $n=2$, 2.5, 3.33, and 5.}\\

\section{Quantum Description}

For a 2D wave-guide system composed of a cavity of arbitrary shape 
connected to two leads, say left (L) and right (R) leads, the solutions in 
the leads are 

\begin{equation}
\Psi^{L,R}(x,y) = \sum_{m=1}\left[a^{L,R}_m \exp(i k^{L,R}_m x) + b^{L,R}_m \exp(-i k^{L,R}_m x)\right]\phi_m(y),
\end{equation}

\noindent where $\phi_m(y)= \sqrt{\frac{2}{d}} \sin \left( \frac{m\pi y}{d} 
\right)$ is the component of the wave function along the $y$-axis. The sum is 
over all the propagating modes $M$ supported by the leads at a given Fermi 
Energy $E$.\

The quantum scattering problem is completely described by the $S$-matrix. It relates incoming waves to outgoing waves, $V^{out} = \hat{\bf S}\ V^{in}$, where $V^{in}$ and $V^{out}$ stand for vectors specifying, 
respectively, waves coming into and going out of the cavity:
\begin{eqnarray}
\hat{\bf S} = \left( \begin{array}{ll} t & r' \\ r & t' \end{array} 
\right ),\ \ V^{in} = \left( \begin{array}{l} a^L \\ b^R \\ \end{array}
\right),\ \ V^{out} = \left( \begin{array}{l} a^R \\ b^L \end{array}
\right).
\end{eqnarray}

Here $t$ ,$t'$, $r$, and $r'$ are the transmission and reflection $M \times M$ matrices.

\section{The Resonance Effect}

We have seen, in the case of global chaos, very good quantum-classical correspondence (QCC) between SP matrices $(\mid S_{n,m} \mid^2)$ \cite{finite1,finite2}, Transient Poincar\`{e} Maps (TPM) \cite{finite2}, and wavefunctions \cite{next}. But this QCC is not generally as good for typical systems with mixed phase space. Moreover, it is well known that for mixed phase space, transport quantities are not smooth functions of the energy. For example the conductance $G=\frac{2e}{\hbar^2}\sum_{n} \sum_{m}\mid t_{n,m} \mid^2$ (where $t_{n,m}$ are the transmission elements of the scattering matrix $\hat{\bf S}$) presents sharp resonances, see Fig. 2. These resonances in $G$ are due to two types of states: regular states, produced by quantum tunneling to classical forbidden regions of phase space; and {\it hierarchical states}, see \cite{ketz} for details. Previously we have studied the QCC for energy values off resonances but for this work the relevant situation occurs for specific resonant values of energy: those related to regular states. As an example, in Fig. 3 wavefunctions $\mid \Psi^L(x,y) \mid^2$ are shown for the resonance energy value of the inset of Fig. 2. In \cite{finite2} it was shown, using Husimi distribution functions \cite{hus} that the maximum probability in phase space of these {\it ``bow-tie''-shaped wavefunctions} is located inside the two islands (forbidden for the scattering problem) centered at $\sin(\chi)=0.3$ in the PM of Fig. 1.\

So, we have seen that certain sharp dips in $G$ are due to ``bow-tie''-shaped resonances in the cavity produced, in turn, by quantum tunneling into classical forbidden regions of phase space (resonance islands).

\begin{figure}[htb]
\begin{center}
\epsfig{file=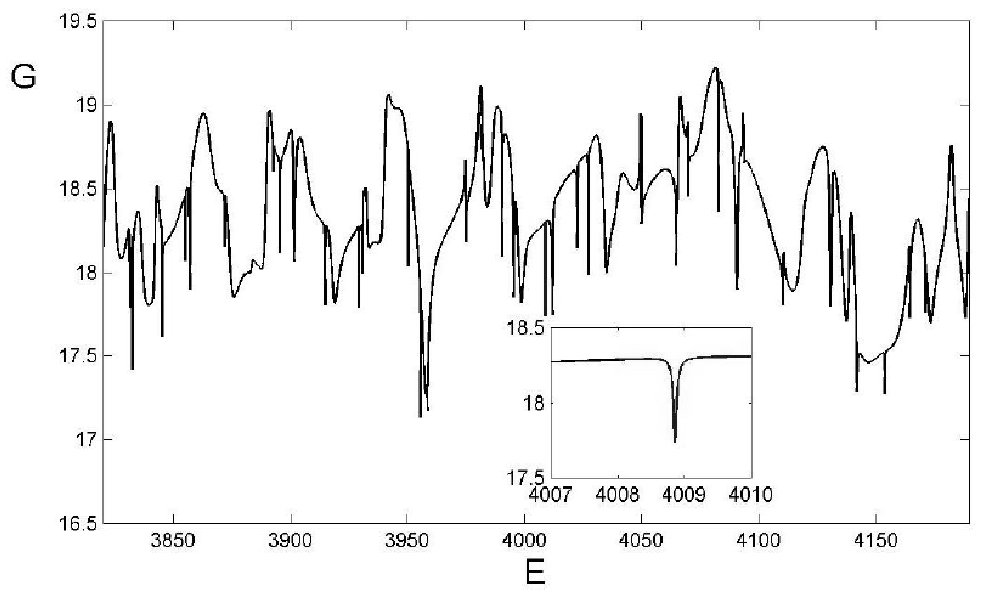,width=5in,height=3.5in}
\label{weaal0}
\end{center}
\end{figure}
\vspace{-0.28in}
{\small FIG. 2 Dimensionless conductance $G$ as a function of the normalized energy $E$. 20 propagating modes are open in this energy range. {\it Inset:} resonance energy for which the wavefunctions have their support inside the two islands centered at $\sin(\chi)=0.3$ in the PM of Fig. 1.}\\

\begin{figure}[htb]
\begin{center}
\epsfig{file=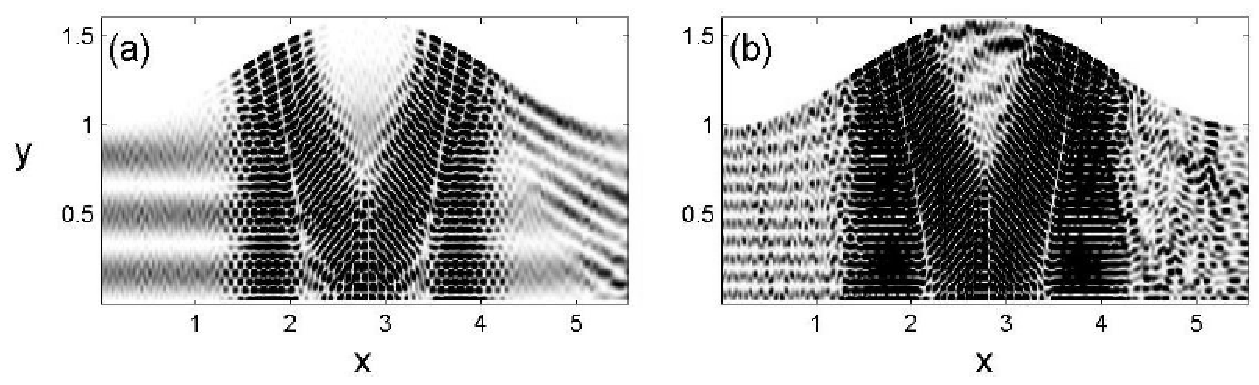,width=6.5in,height=2in}
\label{weaal0}
\end{center}
\end{figure}
\vspace{-0.28in}
{\small FIG. 3 Wavefunctions $\mid \Psi^L(x,y) \mid^2$ calculated for the energy of the inset of Fig. 2 that produces a resonance in $G$. These ``bow-tie''-shaped wavefunctions correspond to modes (a) 3, and (b) 10.}\\

\section{The Open Resonator}

Now we propose to use the wave-guide system described above with an energy 
corresponding to a resonance in $G$ as an optical resonator. If we construct 
this device with a semiconductor having a refraction index $n$ (as described before), then we are able to predict directional lasing emission using the PM.\

From ray optics we know that a ray trajectory inside the cavity will remain inside if the angle of insidence $\chi$ at which it hits the boundary (where the PSS is located) is larger than the critical angle $\chi_c = \sin^{-1}(1/n)$. But when $\chi < \chi_c$ the ray escapes from the semiconductor cavity according to Snell's law. In Fig. 1 we show some critical lines for refractive scape depending on the index of refraction $n$ of the semiconductor used in the construction of the wave-guide. {\it I.e.} rays impinging on the boundary below the critical line will scape from the cavity while the others will reflect.\

Remember that the wavefunctions of Fig. 3 have their main support on the islads centered at $\sin(\chi)=0.3$, then for initial conditions inside these islands and following the ray-optics dynamics of such trajectories, it is possible to have an estimate of the direction and intensity of the lasing emission from the semiconductor cavity depending on the value of $n$. In Fig. 4 we show the patterns obtained for lasing emission using ray-optics dynamics.\

We have considered four values of $n$ in Figs. 1 and 4. When $n = 2$ and 2.5 the islands on which the ``bow-tie'' modes (see Fig. 3) have their support are below the corresponding critical lines for refractive scape. In such situations the emission is similar in intensity but different in direction because $n$ afects the refraction angle according to Snell's law, see Figs. 4a-b. When $n=3.33$ the critical line goes trough the middle of the islands and the intensity of the emited light decreases, as well as the emission angle (see Fig. 4c). For $n=5$ the islands are above the critical line and light cannot scape from the cavity (see Fig. 4d). This is the behavior we expect to observe when testing different semiconductor materials with the appropiate choice of the Fermi Energy. In \cite{arc1,arc2} it has been shown that the ray-optics prediction of lasing emission is in rather good agreement with experimental realizations, so we expect that the predictions in Fig. 4 are also valid.\

Note that it is possible to excite different ``bow-tie''-shaped modes by using different energy values or also by varying the geometry. For example, using the same geometrical parameters $(d,a,L) = (1,0.305,5.55)$ it is possible to find a wavefunction with phase space support inside the big island located at $\sin(\chi)=0$ (see Fig. 1) which will produce perpendicular lasing emission \cite{next}.\\

\begin{figure}[htb]
\begin{center}
\epsfig{file=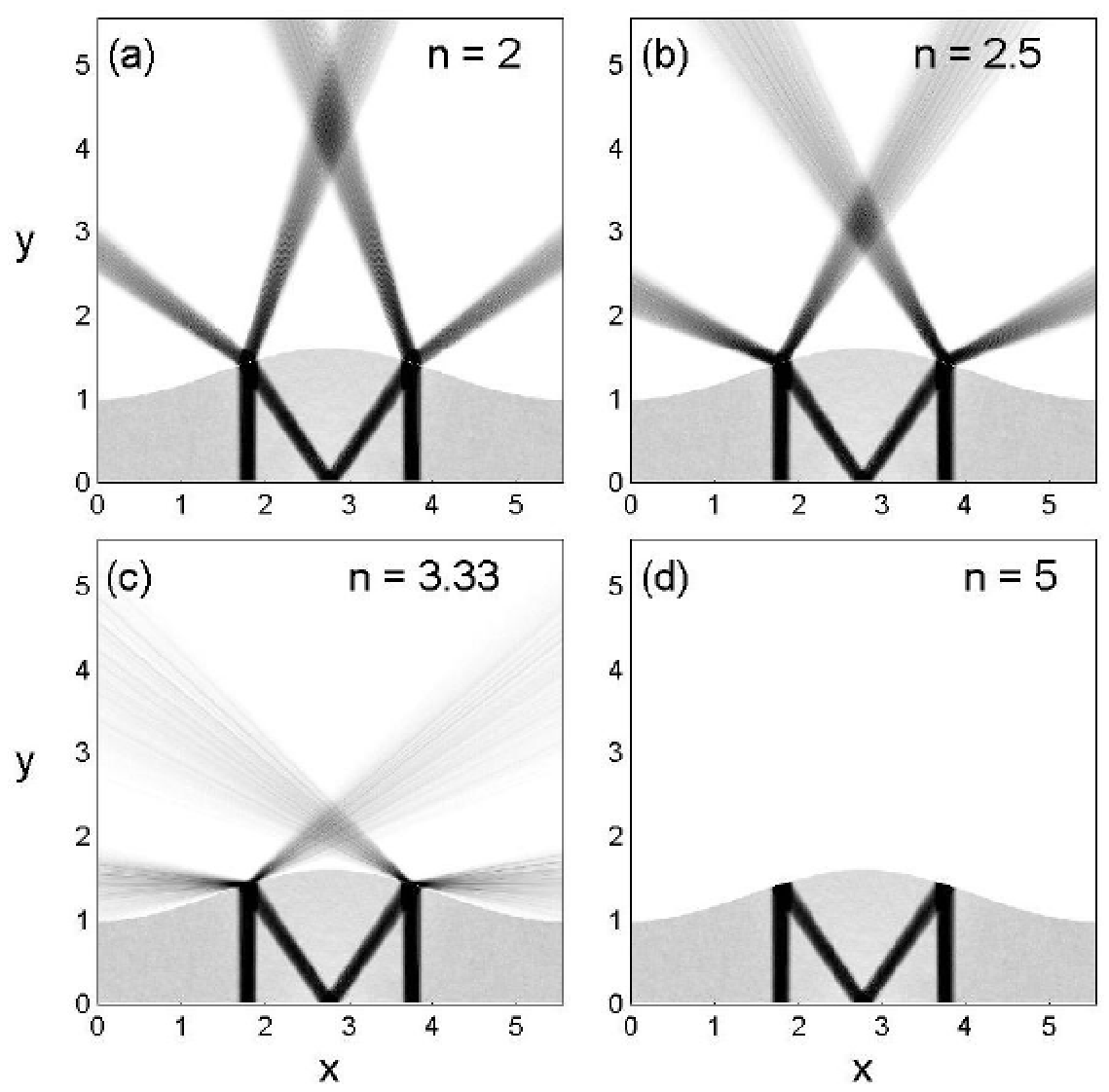,width=5in,height=5in}
\label{weaal0}
\end{center}
\end{figure}
\vspace{-0.28in}
{\small FIG. 4 Ray-optics prediction of lasing emission for $n=2$, 2.5, 3.33, and 5.}

\section{Conclusions}

Here we introduce a prototype of {\it open} semiconductor resonator with controlable directional emission that may be realized experimentally, and may have potential aplications in the design of microlasers and in the field of optical communication systems.

\end{document}